\documentclass{PoS}

\usepackage{amsmath}
\usepackage{amssymb}
\usepackage{graphicx}
\usepackage{comment}
\usepackage{dsfont}

\def\Slash#1{\setbox0=\hbox{$#1$} 
\dimen0=\wd0 
\setbox1=\hbox{/} \dimen1=\wd1 
\ifdim\dimen0>\dimen1 
\rlap{\hbox to \dimen0{\hfil/\hfil}} 
#1 
\else 
\rlap{\hbox to \dimen1{\hfil$#1$\hfil}} 
/ 
\fi}

\title{Faddeev equations: a view of baryon properties }

\ShortTitle{Faddeev equations: a view of baryon properties }

\author{\speaker{D. Nicmorus }\\
        Institut f\"ur Physik, Karl-Franzens-Univert\"at Graz, A-8010 Graz, Austria\\
        E-mail: \email{diana.nicmorus@uni-graz.at}}

\author{G. Eichmann\\ 
        Institut f\"ur Physik, Karl-Franzens-Univert\"at Graz, A-8010 Graz, Austria\\
        E-mail: \email{ge.eichmann@uni-graz.at}}

\author{A. Krassnigg  \\  
         Institut f\"ur Physik, Karl-Franzens-Univert\"at Graz, A-8010 Graz, Austria\\
         E-mail: \email{andreas.krassnigg@uni-graz.at}}

 \author{ R. Alkofer\\      
         Institut f\"ur Physik, Karl-Franzens-Univert\"at Graz, A-8010 Graz, Austria\\
         E-mail:  \email{reinhard.alkofer@uni-graz.at}}
	       
\abstract{ We present a calculation of the three-quark core contribution to the mass of the $\Delta $-resonance 
        in a Poincar\'{e}-covariant Faddeev framework. A consistent setup for the dressed-quark propagator,
        the quark-quark and quark-'diquark' interactions is used, where all the ingredients are solutions of
        their respective Dyson-Schwinger or Bethe-Salpeter equations in rainbow-ladder truncation.
        We discuss the evolution of the $\Delta$ mass with the current-quark mass and compare to the previously
        obtained mass of the nucleon.}

\FullConference{8th Conference Quark Confinement and the Hadron Spectrum\\
                 September 1-6, 2008\\
                 Mainz. Germany}

\begin{document}

\section{Introduction}

        Present advanced experimental facilities offer a precise test of the composite structure of the nucleon. 
        While its basic properties are satisfactorily described within various approaches (e.g.\ quark models, chiral 
        effective field theory methods, lattice QCD), a correct understanding of the rich structure of the nucleon shall 
        be extracted in terms of QCD's quark and gluon degrees of freedom. 
        Dynamical chiral symmetry breaking and confinement, two prominent phenomena tightly connected with 
        the formation of bound states, 
        can be addressed only within a non-perturbative approach to QCD. In this respect         
        Dyson-Schwinger equations (DSEs) represent a 
	continuum tool to access these aspects (for recent reviews see 
        \cite{Fischer:2006ub,Roberts:2007jh}). They form an infinite set of coupled integral equations for the 
        Green functions. Bound states (mesons and baryons) correspond to poles in the respective $n$-point Green function.
         The residue defines the bound state amplitude. It is these amplitudes that contain all the necessary information 
        to achieve a correct theoretical understanding of the hadron properties.      
        Hadron bound-state equations provide these amplitudes: for example, mesons ($q\bar{q}$ bound-states)  have 
        been successfully explored in Bethe-Salpeter equation (BSE) approaches, while baryons (poles in a 6-quark 
        Green function) are studied by means of a covariant equation of the Faddeev type (a quark-'diquark' bound-state 
        BSE describing three-particle kernels). Such an approach to baryons was motivated by the observation that 
        the attractive nature of quark-antiquark 
	correlations in a color-singlet meson is also attractive for $\bar{3}_C$ quark-quark correlations 
	within a color-singlet baryon \cite{Cahill:1987qr,Maris:2002yu}. Presently, the baryon analysis has to some extent reached the level of 
	sophistication in the mesonic sector by fully solving the DS (quark propagator) and BS (diquark amplitudes) 
	equations in a simple consistent setup: a rainbow-ladder truncation which correctly implements chiral symmetry 
	and its spontaneous breaking. 
       The RL truncated DSEs have been widely used with the aim to reproduce hadron observables: masses and electromagnetic properties of mesons and baryons, 
        e.g.\ \cite{Maris:2006ea,Holl:2005vu,Maris:2005tt,Bhagwat:2006pu,Cloet:2008wg,Eichmann:2007nn}. 
        However, evidence has been provided that important corrections to RL truncation 
        are mesonic effects in the quark and quark-gluon-vertex DSEs, see e.g.\ \cite{Fischer:2008wy}, 
        expected to contribute a significant amount of attraction in the 
	chiral limit while decreasing in the heavy-quark limit. 
	Non-resonant terms beyond RL are assumed to  provide further attraction in the pseudoscalar and vector mesons sector
	\cite{Bhagwat:2004hn,Matevosyan:2006bk}. 
	Moreover, contributions associated with an infrared-divergent quark-gluon vertex \cite{Alkofer:2006gz,Alkofer:2008tt} are 
	expected to be small for most hadronic observables.
        Therefore in the present study we employ the recent prescription of \cite{Eichmann:2008ae} and identify RL results with the hadron quark-core 
        contributions (for applications to the $N$ mass and form factors and the $\Delta$ mass, see \cite{Eichmann:2008ae,
        Eichmann:2008ef,Nicmorus:2008vb}).

\section{A Poincar\'{e} covariant approach to baryons}

\begin{figure}[htp]
                    \begin{center}
                    \includegraphics[scale=0.36]{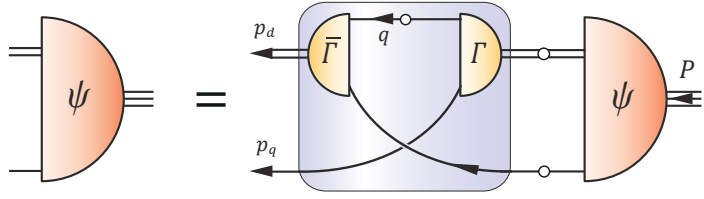}
                    \caption{The quark-diquark BSE . $\Psi$ is the covariant amplitude 
                    for a baryon of total momentum $P$.  $p_q$ and $p_d$ label the dressed-quark and 
                    diquark momenta, respectively. The binding mechanism in the baryons
                    is realized via an iterated exchange of roles between the bystander quark and any of the quarks 
                    contained in the diquark.} \label{fig1}
                    \end{center}
        \label{fig.1}
        \end{figure}

       In a quark-'diquark' scenario, a color-singlet baryon emerges as a bound state of a color-triplet quark and a 
       color-antitriplet diquark correlation. The baryon amplitude and mass are obtained as exact solutions of 
       the baryon quark-diquark BSE 
       (diagramatically represented in Fig.~\ref{fig1}) if all its ingredients are fully specified: 
       the dressed-quark propagator (single line), the diquark propagator (double line), 
       and the diquark amplitude $\Gamma^\nu$ and its charge-conjugate $\bar{\Gamma}^\nu$ which appear 
       in the quark-diquark kernel.
       The renormalized dressed-quark propagator is obtained from its DSE. 
       Its kernel includes the dressed gluon propagator, one bare and one dressed quark-gluon vertex. 
       Upon operating a RL truncation on this kernel, the dressed quark-gluon vertex retains only the vector 
       part $\sim \gamma_\mu$ out of the twelve possible tensors in its general structure, while the dressings 
       of the gluon propagator and of the quark-gluon vertex are absorbed into an effective coupling $\alpha(k^2)$. 
       In this setup the axial-vector Ward Takahashi identity is preserved: the pion remains massless in the chiral 
       limit, while at finite quark masses one obtains a generalized Gell-Mann-Oakes-Renner formula for pseudoscalar 
       mesons \cite{Maris:1997hd,Maris:1997tm}. A convenient choice for the effective coupling is given by \cite{Maris:1999nt} :

                     \begin{equation}\label{couplingMT}
                         \alpha(k^2,\hat{m},\omega) = c(\hat{m}) \,\alpha_{IR}(k^2, \omega)\,+\,\alpha_{UV}(k^2),
                    \end{equation}
        where $k^2$ is the gluon momentum, $\hat{m}$ is the renormalization-point independent current-quark mass. 
        The second term $\alpha_{UV}(k^2)$ decreases logarithmically for large gluon momenta and reproduces 
        QCD's perturbative running coupling. The first term of the interaction is the seed of the model and 
        characterized by a width parameter $\omega$. It accounts for the non-perturbative enhancement at 
        small and intermediate gluon momenta which provides the necessary strength to allow for dynamical 
        chiral symmetry breaking and the dynamical generation of a constituent-quark mass scale. 
        The method of \cite{Eichmann:2008ae} relies upon a parametrization of the BSE solutions of the core $m_\rho(m_\pi)$:
 
            \begin{equation}\label{core:mrho}
                x_\rho^2 = 1 + x_\pi^4/(0.6+x_\pi^2), \;\; x_\rho = m_\rho/m_\rho^0, \;\; x_\pi = m_\pi/m_\rho^0,
            \end{equation}
       with the chiral-limit value $m_\rho^0 = 0.99$ GeV.
       The above parametrization  imposes a 
       dependence of the coupling strength $c(\hat{m})$
        of Eq.~(\ref{couplingMT}) with the current-quark mass determined in \cite{Eichmann:2008ae}. 
        
        In the bound-state equation of Fig.~\ref{fig1} diquarks were implemented as pseudoparticle poles in 
        the quark-quark scattering matrix. Their on-shell amplitudes $\Gamma^{\nu}$ are obtained from 
        the corresponding scalar and axial-vector diquark BSE, whose kernel is (consistently) RL truncated, 
        while their off-shell behavior is parametrized \cite{Eichmann:2008ef} (as imposed by the UV 
        behavior of the two-quark scattering matrix). 
	The nucleon is well described by scalar and 
        axial-vector diquark correlations, while only the axial-vector diquarks contribute to the spin$-\frac{3}{2}$ and 
        isospin$-\frac{3}{2}$ flavor symmetric $\Delta$. 
	
\section{Nucleon and $\Delta$ masses}

        To compute the amplitudes and the masses of the $\Delta$ and N baryons numerically, 
        the on-shell quark-diquark amplitudes $\Psi$ of Fig.~\ref{fig1} are decomposed into an orthogonal set 
        of Dirac covariants (for details see \cite{Eichmann:2008ef,Oettel:1998bk}). The corresponding 
        partial-wave decomposition in terms of eigenfunctions of the quark-diquark total spin and orbital angular 
        momentum in the respective baryon's rest frame reveals an admixture of $p$ and $d-$waves in the nucleon 
        amplitude and $p,d$ and $f-$wave components in the $\Delta$ amplitude, which can be linked to 
        the deviation from sphericity of both nucleon and delta.

\begin{figure}
                    \begin{center}
                    \includegraphics[scale=0.73]{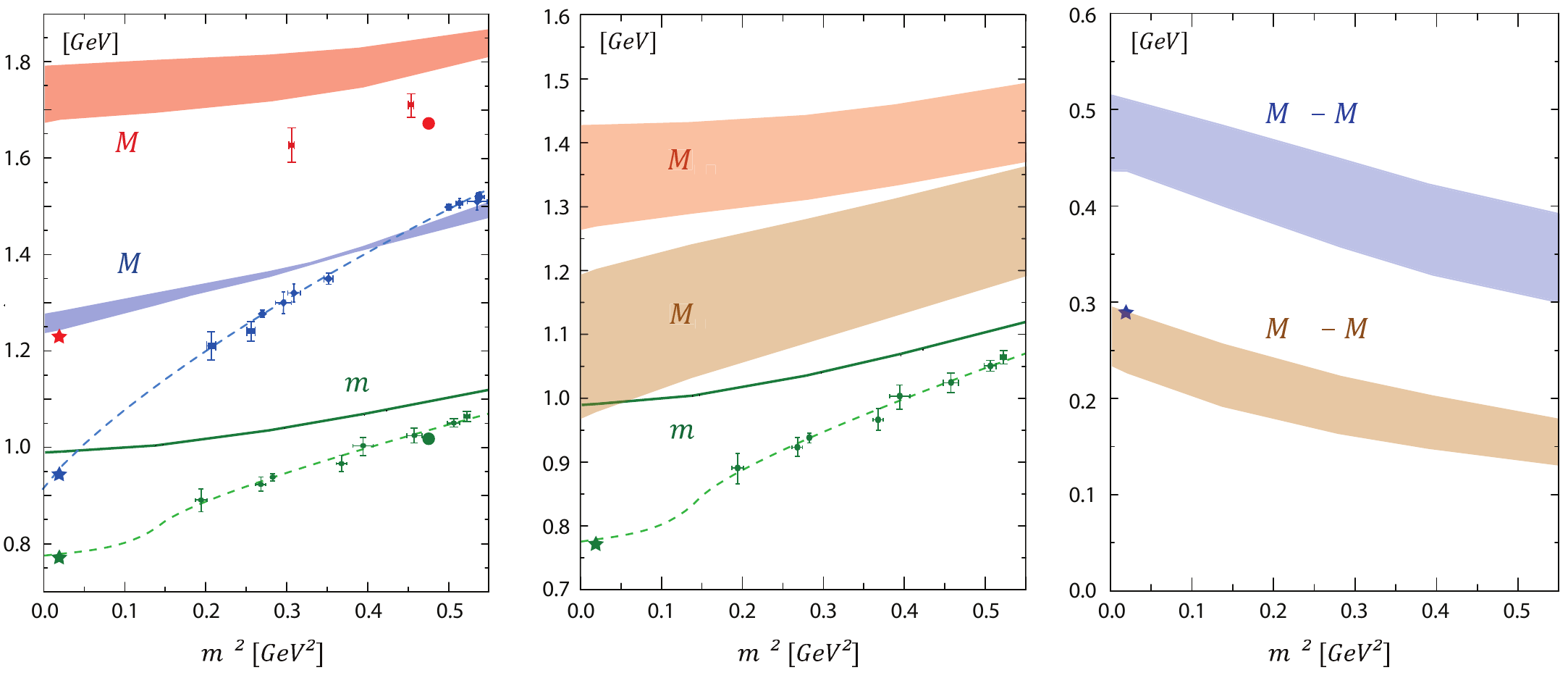}
                    \caption{Evolution of $N$ and $\Delta$ masses (\textit{left panel}), 
		             evolution of $m_\rho$, scalar and axial-vector diquark masses (\textit{center panel}),
                             and the mass splittings $M_\Delta-M_N$ and $M_\textrm{av}-M_\textrm{sc}$ (\textit{right panel})
                             vs. $m_\pi^2$ as obtained from the pseudoscalar $q\bar{q}$ BSE.                                                  
                             We present a selection of lattice data and their chiral extrapolations (\textit{dashed lines}) for
                             $m_\rho$ \cite{AliKhan:2001tx,Allton:2005fb}, $M_N$ \cite{AliKhan:2003cu,Frigori:2007wa,Leinweber:2003dg}
                             and $M_\Delta$ \cite{Zanotti:2003fx}.
                             Experimental values: stars for the $N$ and $\Delta$, filled circles for the $\phi(1020)$ and $\Omega(1672)$.} \label{fig2}
                    \end{center}
        \end{figure}

        The nucleon and $\Delta$ masses emerge upon solving the quark DSE and the diquark and quark-diquark BSE. 
        We depict the results in Fig.~\ref{fig2}. The solid curve for $m_\rho$ represents the chosen core dependence of 
        Eq.\,(\ref{core:mrho}) and fixes the parameters in our interaction.
        The bands represent the results for $M_N$ and $M_\Delta$ and show the sensitivity on the width 
	parameter $\omega$.	At the physical pion mass we obtain $M_\Delta = 1.73(5)$ GeV \cite{Nicmorus:2008vb} and  
	$M_N=1.26(2)$ GeV \cite{Eichmann:2008ef}. We observe that the core model uniformly overestimates these values by $\sim 30 \%$ 
	in the chiral limit. At larger quark masses the deviation from the lattice data decreases, which is in accordance with the core model of 
	Eq.~(\ref{core:mrho}): beyond-RL corrections to most hadronic observables become negligible in the heavy-quark limit. 
	It is evident from Fig.~\ref{fig2} that the experimental values (filled circles) of the $\phi$ meson and the $\Omega$ baryon are 
	overestimated by the same percentage, namely $\sim 8 \%$ (the pseudoscalar value $m_{s\bar{s}}=0.69$ GeV corresponds to the BSE solution at
	 a strange quark mass $\hat{m_s}=150$ MeV \cite{Holl:2004fr}). 
        
        Hadron Dyson-Schwinger studies and chiral analyses of lattice results (e.g.\ \cite{Young:2002cj,Hecht:2002ej}) 
        predict pseudoscalar meson corrections 
        of about $ 200-400$ MeV to the nucleon mass and of a smaller or similar size to the delta mass. 
        Our value for $M_N$ in the chiral region is consistent with a 
	pseudoscalar-meson dressing as dominant correction to the quark-diquark 
	core, but the result for $M_\Delta$ is not. The larger corrections needed in this case could 
	be provided via an improved treatment of the quark-core: inserting further diquark channels, employing the full $qq$ scattering
	kernel instead of the diquark pole or including irreducible 3-body interactions.           
        Also the sizeable $\omega$ dependence of $M_\Delta$ could suggest that taking into account only 
	an axial-vector diquark might not be sufficient to describe the delta baryon, and further diquark components could 
	diminish the $\Delta$ 'core' mass. We observe that although the scalar and axial-vector diquark masses display a significant dependence 
	with $\omega$ (second panel of Fig.~\ref{fig2}), they sensitively cancel when constituting the 
	nucleon mass. Finally, a comparison of the splitting $M_\Delta-M_N$ with the diquark mass splitting $M_\textrm{av}-M_\textrm{sc}$ 
	(third panel Fig.~\ref{fig2}) shows that they both decrease with increasing current-quark mass although there is no 
	direct relationship between the two quantities \cite{Nicmorus:2008vb}.

	We conclude that in the current Poincar\'{e} covariant setup, the RL truncation of the QCD DSEs delivers inflated 'core' contributions. 
	A full understanding of the nucleon structure ultimately relies upon implementing chiral corrections to RL. 
	Their overall attractive effect 
	is expected to shift the core results to the experimental values.

\vspace{-0.4cm}
\section*{Acknowledgements}
    This work has been supported by the Austrian Science Fund FWF under Projects No.~P20592-N16, P20496-N16,
	    and Doctoral Program No.~W1203 as well as by the BMBF grant 06DA267. 
	    D.N. thanks the conveners of Session A, J. Greensite, 
	    M. Polikarpov and M. Faber for their
	    support.

\end{document}